\documentclass[9pt,twocolumn,twoside]{pnas-new}
\templatetype{pnasresearcharticle} % Choose template 
\setboolean{displaywatermark}{false}

\usepackage{graphicx} 
\usepackage{times}
\usepackage{amsmath,amssymb}
\usepackage{color}
\newcommand{\cor}[1]{{#1}}

\title{Experimental observation of the geostrophic turbulence regime of rapidly rotating convection}

\author[a]{Vincent Bouillaut}
\author[a]{Benjamin Miquel} 
\author[b]{Keith Julien}
\author[a]{S\'ebastien Auma\^itre} 
\author[a,1]{Basile Gallet} 

\affil[a]{Universit\'e Paris-Saclay, CNRS, CEA, Service de Physique de l'Etat Condens\'e, 91191 Gif-sur-Yvette, France.}
\affil[b]{Department of Applied Mathematics, University of Colorado, Boulder, Colorado 80309, USA.}

\leadauthor{Bouillaut} 

\significancestatement{Turbulent convection is the main process through which nature moves fluids around, be it in deep planetary and stellar interiors or in the external fluid layers of planets and their satellites.  Laboratory studies aim at reproducing the resulting fully turbulent flows, with the goal of determining the effective transport coefficients to be input into coarse geophysical or astrophysical models. Crucial to these applications is planetary or stellar rotation, which competes with convective processes to set the emergent transport properties. Building on a recent experimental approach that bypasses the limitations of boundary-forced convective flows, we report laboratory measurements in quantitative agreement with the fully turbulent regime of rotating convection.}

\correspondingauthor{\textsuperscript{1}To whom correspondence should be addressed. E-mail: basile.gallet@cea.fr}


\begin{abstract}
The competition between turbulent convection and global rotation in planetary and stellar interiors governs the transport of heat and tracers, as well as magnetic-field generation. These objects operate in dynamical regimes ranging from weakly rotating convection to the `geostrophic turbulence' regime of rapidly rotating convection. However, the latter regime has remained elusive in the laboratory, despite a worldwide effort to design ever-taller rotating convection cells over the last decade.
Building on a recent experimental approach where convection is driven radiatively, we report heat transport measurements in quantitative agreement with this scaling regime, the experimental scaling-law being validated against direct numerical simulations (DNS) of the idealized setup. The scaling exponent from both experiments and DNS agrees well with the geostrophic turbulence prediction. The prefactor of the scaling-law is greater than the one diagnosed in previous idealized numerical studies, pointing to an unexpected sensitivity of the heat transport efficiency to the precise distribution of heat sources and sinks, which greatly varies from planets to stars.
\end{abstract}

\dates{This manuscript was compiled on \today}
\doi{\url{www.pnas.org/cgi/doi/10.1073/pnas.XXXXXXXXXX}}

\begin{document}

\maketitle
\thispagestyle{firststyle}
\ifthenelse{\boolean{shortarticle}}{\ifthenelse{\boolean{singlecolumn}}{\abscontentformatted}{\abscontent}}{}

% If your first paragraph (i.e. with the \dropcap) contains a list environment (quote, quotation, theorem, definition, enumerate, itemize...), the line after the list may have some extra indentation. If this is the case, add \parshape=0 to the end of the list environment.

\dropcap{T}he strong buoyancy gradients inside planets and stars drive turbulent convective flows that are responsible for the efficient transport of heat and tracers, as well as for the generation of the magnetic fields of these objects through the dynamo effect. This thermal and/or compositional driving competes with the global rotation of the astrophysical object: while moderate global rotation only affects the largest flow structures~\cite{Hanasoge,Hanasoge2020,Schumacher}, rapid global rotation greatly impedes radial motion through the action of the Coriolis force, thereby restricting the convective heat transfer~\cite{Stevenson,Julien98}.
Because astrophysical and geophysical flows operate at extreme parameter values, beyond what will ever be achieved in laboratory experiments and numerical simulations, the characterization of these highly complex flows proceeds through the experimental or numerical determination of the constitutive equation, or scaling-law, that relates the turbulent heat flux to the internal temperature gradients. Extrapolating this scaling-law to the extreme parameter values of astrophysical objects sets the effective transport coefficients, the turbulent energy dissipation rate, the mixing efficiency and the power available to induce magnetic field~\cite{Spiegel71,Kraichnan,Stevenson,Ahlers,Christensen,ChristensenAubert,aurnouPEPI12}. 

\begin{figure*}
\centering
\includegraphics[width=11 cm]{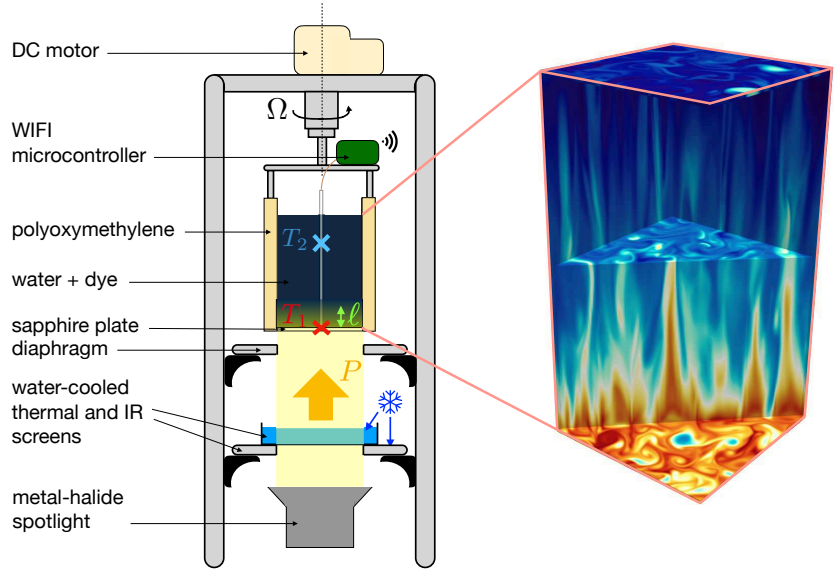} 
\caption{\textbf{Radiatively driven rotating convection.} A powerful spotlight shines from below at a mixture of water and dye. The resulting internal heat source decreases exponentially with height over the absorption length $\ell$, delivering a total heat flux $P$. The cylindrical tank is attached from above to a DC motor that imposes global rotation at a rate $\Omega$ (slight curvature of the top free surface not represented). Two thermocouples $T_1$ and $T_2$ measure the vertical temperature drop in the rotating frame, the data being communicated through WIFI to a remote Arduino microcontroller. On the right-hand side is a DNS snapshot of the temperature field in horizontally periodic geometry devoid of centrifugal and sidewall effects, highlighting the vertically elongated structures of rotating convection ($\mathrm{Ra_P}=10^{12}$, $\text{E}=2 \times 10^{-6}$, $\text{Pr}=7$, $\ell/H=0.048$, arbitrary color scale ranging from blue for cool fluid to red for warm fluid). \label{fig:setup}}
\end{figure*}

Within the Boussinesq approximation~\cite{Spiegel1960} and adopting a local Cartesian geometry, the scaling-laws are cast in terms of the dimensionless parameters that govern the system: the flux-based Rayleigh number $\mathrm{Ra_P}=\alpha g P H^4 / \rho C \kappa^2 \nu$ quantifies the strength of the heat flux $P$, where $H$ denotes the height of the fluid domain, $\alpha$ the coefficient of thermal expansion, $g$ the acceleration of gravity, $\kappa$ the thermal diffusivity, $\nu$ the kinematic viscosity, $\rho$ the mean density and $C$ the specific heat capacity. The Nusselt number $\mathrm{Nu}=PH/\rho C \kappa \Delta T$ measures the heat transport efficiency of the turbulent flow, as compared to that of a steady motionless fluid, in terms of the typical vertical temperature drop $\Delta T$. Finally, the magnitude of the Coriolis force can be quantified through the Ekman number $\mathrm{E}=\nu/2\Omega H^2$, a low value of $\mathrm{E}$ corresponding to a rapid global rotation rate $\Omega$.

At the theoretical level, several arguments have been put forward to predict the scaling-law for the heat transport efficiency of rotationally constrained turbulent convection, as measured by the Nusselt number $\mathrm{Nu}$. Central to these theories is the assumption that the scaling relation between the turbulent heat flux and the internal temperature gradient should not involve the tiny molecular diffusivities $\kappa$ and $\nu$. In the physics community, this assumption is sometimes referred to as the existence of an `ultimate regime' \cite{Chavanne}, while in the astrophysical community it is often referred to as the `mixing-length' regime, because the latter theory neglects molecular diffusivities at the outset~\cite{Vitense,Spiegel71}.

The second assumption is that the heat transport efficiency of the flow depends only on the supercriticality of the system, i.e., on the ratio of the Rayleigh number to the threshold Rayleigh number for the emergence of thermal convection. This idea is put on firm analytical footing through careful asymptotic expansions of the equations of thermal convection in the rapidly rotating limit~\cite{Julien98,Sprague2006,Julien2012,AurnouPRR}. When combined, these two assumptions lead to the following scaling-law for turbulent heat transport by rapidly rotating thermal convection (see Ref.~\cite{Stevenson} for the initial derivation):
\begin{equation}
\mathrm{Nu} = {\cal C} \times \mathrm{Ra_P}^{3/5} \, \mathrm{E}^{4/5} \, \mathrm{Pr}^{-1/5} \, ,\label{scaling32}
\end{equation}
where $\mathrm{Pr}=\nu/\kappa$ is the Prandtl number and ${\cal C}$ is a dimensionless prefactor. Equation [\ref{scaling32}] is referred to as the `geostrophic turbulence' scaling-law of rapidly rotating convection\footnote{Geostrophy refers to the large-scale balance between the Coriolis and pressure forces.}. In terms of the temperature-based Rayleigh number $\mathrm{Ra}=\mathrm{Ra_P}/\mathrm{Nu}$, this scaling-law becomes:
\begin{equation}
\mathrm{Nu} =  {\cal C}_\mathrm{Ra} \times \mathrm{Ra}^{3/2} \, \mathrm{E}^2 \, \mathrm{Pr}^{-1/2} \, , \label{scalingRa}
\end{equation}
where the dimensionless prefactor is ${\cal C}_\mathrm{Ra}={\cal C}^{5/2}$. Over the last decade, several state-of-the-art laboratory experiments have been developed to observe this extreme scaling regime and validate the geostrophic turbulence scaling-law~[\ref{scaling32}]: the TROCONVEX experiment in Eindhoven~\cite{Cheng2020}, the rotating U-boot experiment in G\"ottingen~\cite{Zhang,Wedi2021}, the Trieste experiment at ICTP~\cite{Niemela,Ecke}, and the Romag and Nomag experiments at UCLA~\cite{ChengGJI,KingPNAS}. The goal is to produce a strongly turbulent convective flow in which rotational effects remain predominant (hence the ever taller convective cells), while avoiding parasitic centrifugal effects~\cite{Horn}. These experiments are all based on the Rayleigh-B\'enard (RB) geometry, where a layer of fluid is contained between a hot bottom plate and a cold top one. A particularly challenging task then is to overcome the throttling effect of the boundary layers near these two plates: fluid hardly moves there and heat need be diffused away from those regions~\cite{Malkus}. Even though asymptotic analysis indicates that heat transport should be controlled by the bulk turbulent flow in rapidly rotating RB convection, laboratory realizations indicate that the boundary processes keep limiting the heat transfer throughout the entire cell~\cite{King2009}, bringing the molecular diffusivities back into play and preventing the observation of the scaling-law [\ref{scaling32}] associated with the bulk rotating turbulent flow. Forty years after its initial derivation~\cite{Stevenson} and despite a worldwide effort to design ever taller convection cells, the geostrophic regime of rapidly rotating convection still awaits experimental validation~\cite{Cheng2018}.

%\begin{figure}
%%\hspace{5cm}
%    \centering{\includegraphics[width=8 cm]{Figure1.eps} }
%   \caption{\textbf{Radiatively driven rotating convection.} A powerful spotlight shines from below at a mixture of water and dye. The resulting internal heat source decreases exponentially with height over the absorption length $\ell$, delivering a total heat flux $P$. The cylindrical tank is attached from above to a DC motor that imposes global rotation at a rate $\Omega$ (slight curvature of the top free surface not represented). Two thermocouples $T_1$ and $T_2$ measure the vertical temperature drop in the rotating frame, the data being communicated through WIFI to a remote Arduino microcontroller. On the right-hand side is a DNS snapshot of the temperature field in horizontally periodic geometry devoid of centrifugal and sidewall effects, highlighting the vertically elongated structures of rotating convection ($\mathrm{Ra_P}=10^{12}$, $\text{E}=2 \times 10^{-6}$, $\text{Pr}=7$, $\ell/H=0.048$, arbitrary color scale ranging from blue for cool fluid to red for warm fluid). \label{fig:setup}}
%\end{figure}

%\begin{SCfigure*}[\sidecaptionrelwidth][t]

Recently, we introduced an innovative laboratory setup to overcome the above-mentioned limitations of RB convection as a model for bulk natural flows~\cite{Lepot}. Specifically, we used a combination of radiative internal heating and effective internal cooling to bypass the throttling boundary layers of traditional RB convection and achieve the fully turbulent -- or `ultimate' -- regime of  non-rotating convection~\cite{Lepot,Bouillaut,Miquel2020}. These recent experimental developments suggest an alternative route to observe the geostrophic regime of rapidly rotating turbulent convection in the laboratory: instead of trying to overcome the throttling effect of the RB boundary layers through intense thermal forcing, one can take advantage of the radiatively driven setup, where these boundary layers are readily bypassed, and subject the radiatively driven turbulent convective flow to rapid global rotation.

%\section{Experimental setup}

\begin{figure}[ht]
   % \centerline{\includegraphics[width=13 cm]{figures/Nu_vs_E_raw_article.eps} }
        \centerline{\includegraphics[width=9 cm]{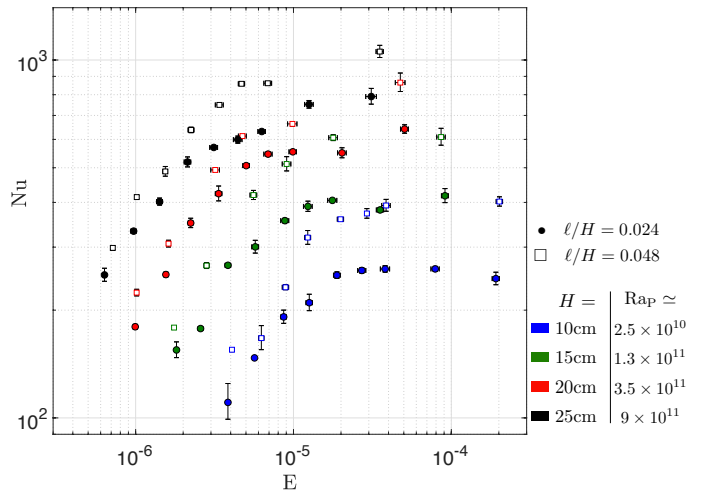} }
   \caption{{\bf Suppression of heat transport by global rotation.} Heat transport efficiency $\mathrm{Nu}$ as a function of the Ekman number $\mathrm{E}$, for various fluid heights: blue, $H=10$cm, $\mathrm{Ra_P}\simeq 2.5 \times 10^{10}$; green, $H=15$cm,  $\mathrm{Ra_P}\simeq 1.3 \times 10^{11}$; red, $H=20$cm,  $\mathrm{Ra_P}\simeq 3.5 \times 10^{11}$; black, $H=25$cm,  $\mathrm{Ra_P}\simeq 9\times10^{11}$. The dimensionless absorption length is $\ell/H=0.024$ (filled circles) or $\ell/H=0.048$ (open squares). For fixed $H$ and $\ell$, the mixing efficiency dramatically decreases with increasing rotation rate (decreasing $\mathrm{E}$). \label{fig:rawdata} Errorbars are estimated from the values obtained for the first and second halves of the measurement interval, see Methods and SI appendix.}
\end{figure}

The resulting experimental setup, sketched in Figure~\ref{fig:setup}, is an evolution over the non-rotating setup described in a previous publication~\cite{Lepot}. The apparatus consists of a cylindrical tank of radius $10$cm with a transparent sapphire bottom boundary, filled with a light-absorbing mixture of water and carbon-black dye. A powerful spotlight located under a water-cooled IR-screening stage shines at the tank from below. Absorption of light by the dye results in an internal heat source that decreases exponentially with height $z$ measured upwards from the bottom of the tank, transferring to the fluid a total heat flux $P$ over an e-folding absorption length $\ell$. This source term causes the temperature at every location inside the tank to increase linearly with time. Superposed to this linear drift are internal temperature gradients that develop inside the tank and rapidly reach a statistically steady state. As recalled in the Methods section, the internal temperature difference between any two points inside the tank is then governed by a combination of the exponential radiative heat source together with an effective uniform heat sink.

The experimental tank is attached from above to a DC motor that drives global rotation at a constant rate $\Omega \in [0;85]$~rpm around the vertical axis of the cylinder. \cor{Rotation results in a slight curvature of the free surface: the relative variations in fluid height between center and periphery reach $\pm 20 \%$ and $\pm 13 \%$ for the two most rapidly rotating and shallowest data points, but are below $\pm 10\%$ (and often much below) for the remaining approximately $60$ data points.} On-board temperature measurements are performed using two thermocouples, one in contact with the bottom sapphire plate and one at $z=3H/4$, where $H$ denotes the height of the free surface on the axis of the cylindrical cell, where the probes are located. The temperature signals are transmitted through WIFI to a remote Arduino microcontroller to ensure live monitoring of the experimental runs.

We show in Figure~\ref{fig:rawdata} the Nusselt number based on the time-averaged temperature difference $\Delta T$ between the two probes, for experimental runs spanning 1.5 decades in $\mathrm{Ra_P}$ and 2.5 decades in $\mathrm{E}$, and two values of the dimensionless absorption length $\ell/H$. The dataset is provided in the Supplemental Information (SI) appendix, together with estimates of the error bars.
In a similar fashion to the more standard RB system, for an approximately constant $\mathrm{Ra_P}$\footnote{As shown in the SI table, the temperature range varies between different data points, the consequence being that $\mathrm{Ra_P}$ and $\mathrm{Pr}$ vary between different points of a constant-$H$ curve in Figure~\ref{fig:rawdata}. The entire range of $\mathrm{Pr}$ spanned by the experimental data is $4.4 \leq \mathrm{Pr} \leq 6.7$.} an increase in the global rotation rate leads to a dramatic drop in the heat transport efficiency as measured by the Nusselt number $\mathrm{Nu}$. 

%

%\section{Geostrophic turbulence regime of rapidly rotating convection}

\begin{figure}
%    \centerline{\includegraphics[width=14 cm]{figures/rescaledNu_vs_controlparam_article.eps} }
\vspace{-3cm}
    \centerline{\includegraphics[width=9 cm]{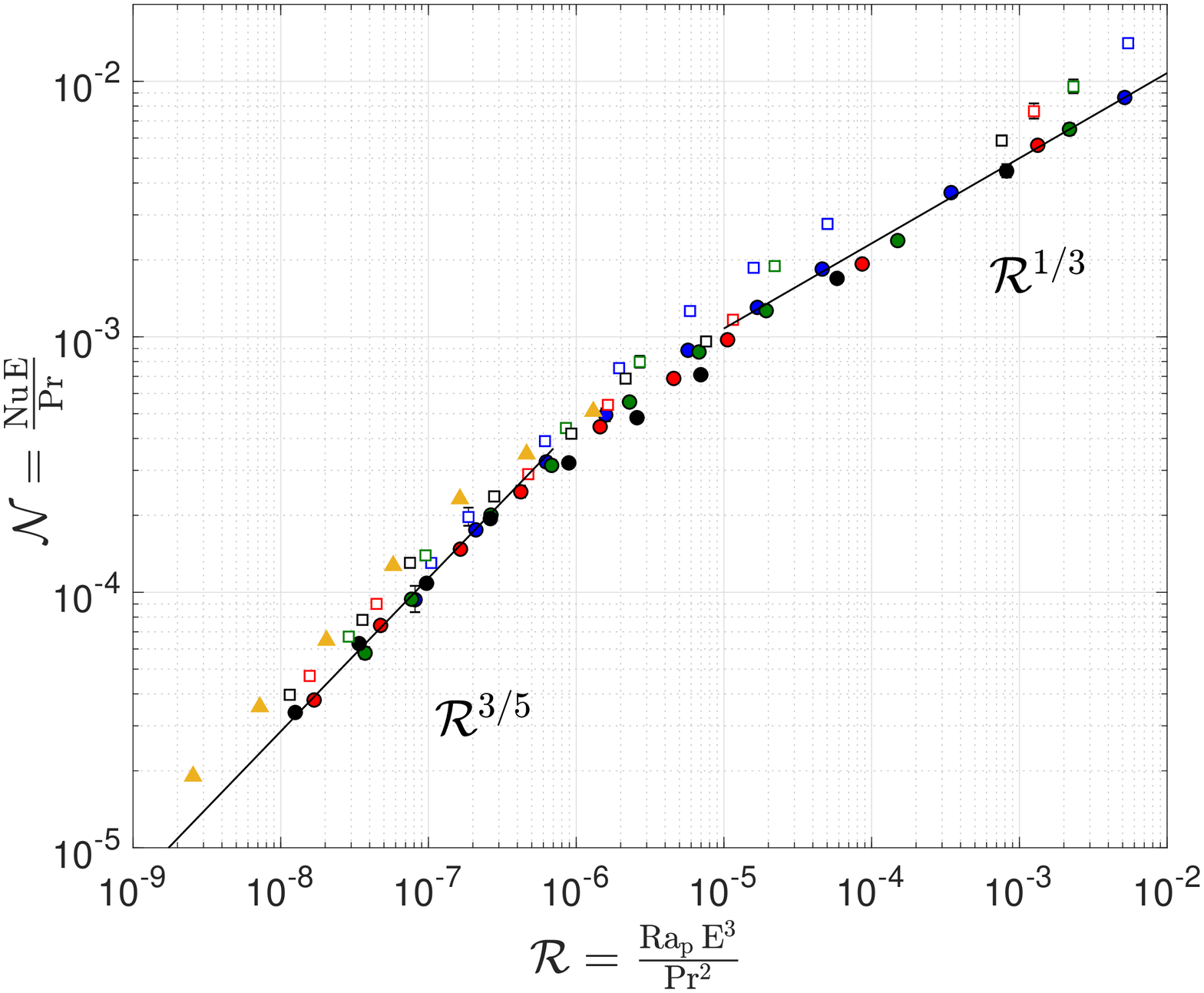} }
        \centerline{\includegraphics[width=9 cm]{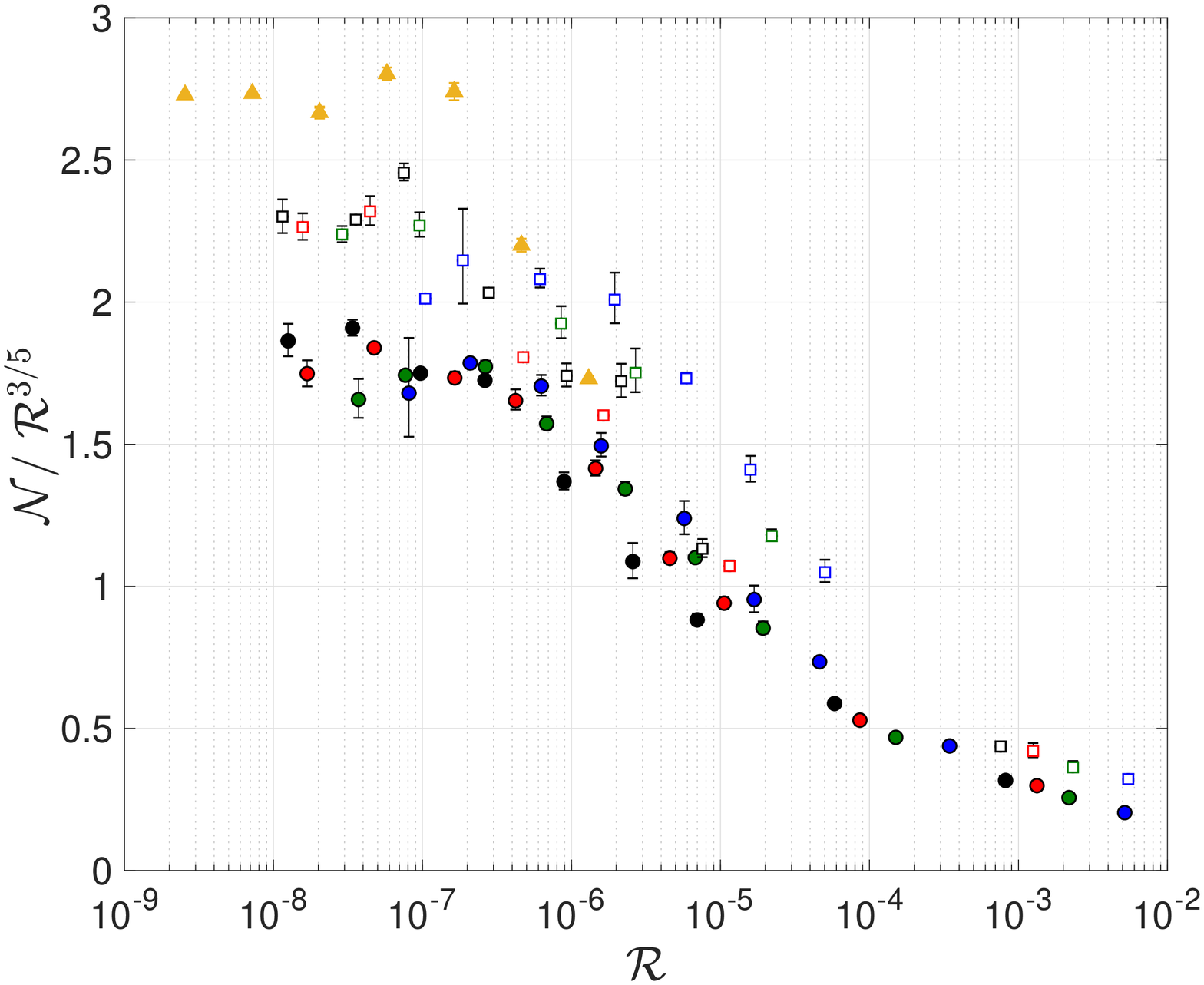} }
   \caption{{\bf Observation of the geostrophic turbulence regime.} In terms of the diffusivity-independent parameters ${\cal N}$ and ${\cal R}$, the data gathered for a given value of $\ell/H$ collapse onto a master curve, which validates the `fully turbulent' assumption. In the rapidly rotating regime ${\cal R} \lesssim 3\times 10^{-7}$ the master curve displays a power-law behavior over one and a half decades in ${\cal R}$, in excellent agreement with the prediction ${\cal N}\sim {\cal R}^{3/5}$ associated with the geostrophic turbulence scaling regime of rapidly rotating convection (shown as an eye-guide, see Table~\ref{table:beta} for best-fit exponents). Same symbols as in Figure~\ref{fig:rawdata} for the experimental data. The triangles are DNS data for $\mathrm{Ra_P}=10^{12}$, $\mathrm{Pr}=7$ and $\ell/H=0.048$. Experimental and numerical error bars are visible when larger than the symbol size. Bottom: same data compensated by the geostrophic turbulence scaling prediction. An approximate plateau is observed for ${\cal R} \lesssim 3\times 10^{-7}$. \label{fig:rescaleddata}}
\end{figure}

With the goal of establishing the turbulent nature of the flow and assessing the independence of its transport properties with respect to the molecular diffusivities, we form the $\nu$- and $\kappa$-independent reduced Nusselt number ${\cal N}=\mathrm{Nu} \, \mathrm{E}/\mathrm{Pr}$, together with the composite control parameter ${\cal R}=\mathrm{Ra_P} \, \mathrm{E}^3 / \mathrm{Pr}^2$. The latter combination is the only dimensionless control parameter if the diffusivities are to play no roles~\cite{Christensen,ChristensenAubert,Schmitz2009}. ${\cal R}$ is also the cube of the so-called flux-based convective Rossby number, identified as the main control parameter of open ocean convection~\cite{Marshall,AurnouPRR}. We plot ${\cal N}$ as a function of ${\cal R}$ in Figure~\ref{fig:rescaleddata} (data points and estimates of the error provided in the SI). In this representation, the dataset for a given value of $\ell/H$ collapses onto a single master curve, which validates the fact that the molecular diffusivities are irrelevant: we conclude that the present experimental setup achieves a `fully turbulent' scaling regime, according to the definition given at the outset. The collapse is particularly good for rapid global rotation and slow global rotation -- low and large ${\cal R}$, respectively -- with a bit more scatter for intermediate values. For slow global rotation (large ${\cal R}$) the master curve gradually approaches the scaling-law of radiatively driven non-rotating convection, reported in previous publications~\cite{Lepot,Bouillaut,Miquel2020}. This regime is associated with a large-${\cal R}$ asymptote of the form ${\cal N} \sim {\cal R}^{1/3}$, represented in Figure~\ref{fig:rescaleddata}: after crossing out $\mathrm{E}$ from both sides of the scaling relation ${\cal N} \sim {\cal R}^{1/3}$ one recovers the ultimate scaling-law of non-rotating convection, where $\mathrm{Nu}$ is proportional to the square-root of $\mathrm{Ra}$~\cite{Lepot,Bouillaut,Miquel2020}. The approach to that asymptotic behavior is clearly visible for $\ell/H=0.024$ at large ${\cal R}$, with a bit more scatter for $\ell/H=0.048$\footnote{One would probably need to reach larger ${\cal R}$ to avoid any signature of the intermediate-${\cal R}$ scatter.}. More interestingly, the focus of the present study is on the rapidly rotating regime that arises for ${\cal R} \lesssim 3 \times 10^{-7}$. In this parameter range the master curve follows a power-law behavior ${\cal N} \sim {\cal R}^{\beta}$ over one and a half decades. The best-fit exponents $\beta$ are given in Table~\ref{table:beta}. Over the last decade in ${\cal R}$, \cor{we measure $\beta=0.57\pm0.03$ and $\beta=0.62\pm 0.01$}, respectively, for $\ell/H=0.024$ and $\ell/H=0.048$. These values are within $5\%$ of the theoretical exponent $\beta=3/5$ associated with the geostrophic turbulence scaling-law [\ref{scaling32}]. 

While the flux-based parameter ${\cal R}$ is the natural control parameter of the present experiment, the reader accustomed to the standard RB setup may be interested in characterizing the data in terms of the Rayleigh number $\text{Ra}$ based on the emergent temperature gradient. In the SI appendix, we thus plot ${\cal N}$ as a function of the diffusivity-free Rayleigh number $\text{Ra}_*=\text{Ra} \,  \text{E}^2 / \text{Pr}$ (also known as the square of the temperature-based convective Rossby number~\cite{AurnouPRR}). This representation is equivalent to the one in Figure~\ref{fig:rescaleddata}, with an equally satisfactory collapse of the dataset. The power-law fits reported in Table~\ref{table:beta} translate into power-laws ${\cal N}\sim \text{Ra}_*^\gamma$, where the exponent $\gamma$ is within $12\%$ of the theoretical prediction $3/2$ (\cor{$\gamma=1.33\pm 0.14$ and $\gamma=1.63 \pm 0.07$}, respectively, for $\ell/H=0.024$ and $\ell/H=0.048$). %These values are in stark contrast with the very large scaling exponent $\gamma$ in the scaling-law $\text{Nu}\sim \text{Ra}^\gamma$ of rapidly rotating RB convection at constant $E \ll 1$. 

\cor{These values contrast with the scaling exponent $\gamma$ in the constant-$\text{E}$ scaling-law $\text{Nu}\sim \text{Ra}^\gamma$ reported in laboratory studies of rotating RB convection (see Ref.~\cite{Kunnen2021} for a recent review). According to the literature, the RB exponent measured experimentally achieves a value close to $1/3$ in the slowly rotating regime, in line with the `classical theory' of non-rotating RB convection~\cite{Malkus}. For fast rotation and moderate supercriticality, laboratory experiments typically enter a transitional regime where the exponent $\gamma$ increases sharply. An extension of the experimental data using Direct Numerical Simulations (DNS) indicates that $\gamma$ eventually reaches a value ranging between $3$ and $4$~\cite{ChengGJI}, the lower value $3$ being again associated with a `classical' regime controlled by marginally stable boundary layers~\cite{King2012}, while the larger value $4$ has been attributed to Ekman pumping~\cite{Plumley2016} (see also Ref.~\cite{Julien2016} for a theoretical demonstration of increasing heat transport exponents as a result of boundary layer pumping).} By contrast, in the present experiment radiative heating bypasses the boundary layers of standard rotating RB convection, thus circumventing the limitations of this traditional setup and providing experimental observations in excellent agreement with the geostrophic scaling regime of rapidly rotating turbulent convection. 

\cor{As a side note, we stress the fact that the system operates far above the instability threshold. In the rapidly rotating limit, convection arises above a threshold value of the order of $15$ for the reduced flux-based Rayleigh number $\mathrm{Ra_P} \, \text{E}^{4/3}$. We report the values of $\mathrm{Ra_P} \, \text{E}^{4/3}$ in the SI appendix: in the rapidly rotating regime this parameter ranges between $1.5\times 10^3$ and $2.5 \times 10^4$, orders of magnitude above its threshold value. This large distance from threshold is confirmed by the large values of the Nusselt number in Figure~\ref{fig:rawdata}, which range between $10^2$ and $10^3$. The collapse in Figure~\ref{fig:rescaleddata} is thus not a mere consequence of near-onset behavior, a phenomenon reported in Cheng \& Aurnou~\cite{Cheng2016} for synthetic near-onset data. We illustrate this point further in the bottom panel of Figure~\ref{fig:rescaleddata}, where we plot ${\cal N}$ compensated by the geostrophic turbulence scaling-law ${\cal R}^{3/5}$, in semi-logarithmic coordinates. This rather stringent representation confirms (i) the good collapse of the data and (ii) the existence of a plateau at low ${\cal R}$, in agreement with the geostrophic turbulence scaling-law. By contrast, the near-onset data discussed by Cheng \& Aurnou would not display such a collapse onto a plateau in compensated form, as illustrated in Figure~S2 of the SI appendix. We further emphasize this point in the SI appendix by plotting the Nusselt number mutliplied by $\sqrt{\text{Pr}}$ -- to collapse the various $\text{Pr}$ data points, according to the geostrophic turbulence scaling -- as a function of the reduced temperature-based Rayleigh number $\text{Ra} \, \text{E}^{4/3}$. The resulting Figure~S3 makes it clear that the present data do not correspond to the near-onset behavior discussed by Cheng \& Aurnou in the RB context: they are associated with greater supercriticality -- the latter being better estimated by the even greater reduced flux-based Rayleigh number $\mathrm{Ra_P} \, \text{E}^{4/3}$ in the present context -- and a scaling exponent compatible with the  $\gamma=3/2$ geostrophic turbulence value (see SI appendix for details).}

%These values contrast sharply with the very large scaling exponent $\gamma$ in the constant-$\text{E}$ scaling-law $\text{Nu}\sim \text{Ra}^\gamma$ reported in laboratory studies of rapidly rotating RB convection. The RB exponent measured experimentally ranges between $3$ and $4$~\cite{ChengGJI}, where the lower value $3$ is associated with a `classical' regime controlled by marginally stable boundary layers~\cite{Malkus,King2012}, while the larger value $4$ has been attributed to Ekman pumping~\cite{Plumley2016}.
%By contrast, in the present experiment radiative heating bypasses the boundary layers of standard rotating RB convection, thus circumventing the limitations of this traditional setup and providing experimental observations in excellent agreement with the geostrophic scaling regime of rapidly rotating turbulent convection.

\begin{table}
\begin{center}
\begin{tabular}{ |c|c|c| } 
 \hline
  $\beta$ & ${\cal R} \leq 3 \times 10^{-7}$ & ${\cal R} \leq 10^{-7}$ \\ 
 \hline
% Experiments $\ell/H=0.024$ &  $0.587 \pm 0.013$  &  $0.572 \pm 0.026$ \\ 
% Experiments $\ell/H=0.048$ &  $0.567 \pm 0.010$   &  $0.616  \pm 0.011$ \\ 
% DNS $\ell/H=0.048$ &  $0.601 \pm 0.002$ &  $0.601 \pm 0.002$ \\
 Experiments $\ell/H=0.024$ &  $0.59 \pm 0.01$  &  $0.57 \pm 0.03$ \\ 
 Experiments $\ell/H=0.048$ &  $0.57 \pm 0.01$   &  $0.62  \pm 0.01$ \\ 
 DNS $\ell/H=0.048$ &  $0.601 \pm 0.002$ &  $0.601 \pm 0.002$ \\
\hline
\end{tabular}
\end{center}
\caption{Best-fit exponent $\beta$ for laboratory and DNS data, to be compared to the theoretical prediction $3/5$ associated with the geostrophic turbulence scaling regime. \cor{The error $\pm \sigma_\beta$ is estimated by propagating the error on $\log {\cal N}$ into a standard deviation $\sigma_\beta$ for the exponent.} \label{table:beta}}
\end{table}

It proves insightful to compare the present experimental results to existing numerical studies. DNS have been used as an extremely valuable tool both to address rotating convection inside full or partial spheres~\cite{Gastine}, but also to develop thought experiments in which one can alter the exact equations and/or boundary conditions to identify the mechanisms at play. This leads to idealized situations in which the geostrophic scaling regime emerges. Some studies have considered stress-free boundary conditions instead of the no-slip boundaries of experimental tanks~\cite{Stellmach,Kunnen}, some have used tailored internal heat sources and sinks that conveniently vanish at the boundaries of the domain~\cite{Barker,Currie}, and some have focused on reduced sets of equations obtained through an asymptotic expansion of the rapidly rotating Boussinesq equations~\cite{guervillyNAT19,Sprague2006,Julien2012}. DNS also offer an opportunity to eliminate the potential biases of laboratory experiments. One source of experimental bias is the centrifugal acceleration, which increases with global rotation rate and distance from the rotation axis. Horn and Aurnou~\cite{Horn} proposed the criterion $\Omega^2 H/g \lesssim 1$ for centrifugal effects to be negligible in standard rotating RB convection (see also Refs.~\cite{Horn2019,Horn2021}). The precise threshold value on the right-hand side of this inequality can probably be debated and requires further investigation, specifically Cheng et al.~\cite{Cheng2020} report experimental measurements unaffected by centrifugal effects even when $\Omega^2 H/g$ is as large as two.  In the SI appendix, we provide the value of $\Omega^2 H/g$ for all the experimental data points: this ratio never exceeds two, with only three data points for which this ratio exceeds one (per value of $\ell/H$). A second distinction between the idealized horizontally unbounded convective layer and the finite-size experimental tanks is  the possible emergence of localized convective modes near the vertical walls of the latter~\cite{Buell1983,Zhong1991,Ecke1992,Favier2020}. Wall modes have a lower onset than bulk modes and can dominate the dynamics near the instability threshold of the latter. However, they have also been shown to have a negligible impact on bulk heat transport in the turbulent regime~\cite{deWit2020}. The collapse of the various data points in Figure~\ref{fig:rescaleddata} -- which differ both in terms of centrifugal ratio $\Omega^2 H/g$ and aspect ratio -- is a first indication that the present measurements are not impacted by the centrifugal acceleration nor the sidewalls.

With the goal of further validating the experimental results and the subdominance of centrifugal, sidewall and non-Boussinesq effects, we have performed DNS of the present radiatively driven setup in the idealized horizontally periodic plane layer geometry. The combination of radiative heating and uniform internal cooling is implemented in a pseudo-spectral code that solves the rotating Boussinesq equations of thermal convection with a no-slip insulating bottom boundary and a stress-free insulating top one (see Methods for details). We provide a snapshot of the temperature field in statistically steady state in Figure~\ref{fig:setup}, for $\mathrm{Ra_P}=10^{12}$, $\text{E}=2 \times 10^{-6}$, $\text{Pr}=7$ and $\ell/H=0.048$. This temperature field displays the typical vertically elongated structures that characterize rapidly rotating convection~\cite{Cheng2018}, with a predominance of thin warm plumes emanating from the heating region. The full numerical dataset consists in a sweep of the Ekman number $\text{E}$ for $\mathrm{Ra_P}=10^{12}$, $\text{Pr}=7$ and $\ell/H=0.048$, the resulting data points being plotted in Figure~\ref{fig:rescaleddata}. The error bars on these data points, provided in the SI appendix, are much smaller than the size of the symbols. The low-${\cal R}$ numerical data again display a power-law behavior, with a best-fit exponent $\beta$ within $0.5\%$ of the theoretical exponent $3/5$ associated with the geostrophic turbulence scaling regime, see Table~\ref{table:beta}. 
The numerical data points lie very close to the experimental ones for the same value of $\ell/H$, the reduced Nusselt number being slightly larger for the DNS data (by approximately 20\%), possibly as a consequence of the somewhat different geometries of the numerical and experimental setups. Overall, the quantitative agreement between experiments and DNS\cor{, together with the good collapse in Figure~\ref{fig:rescaleddata} of data points obtained for various aspect ratios and centrifugal ratios, indicates that the aforementioned potential biases are subdominant in the present experiment}. As far as the present heat transport measurements are concerned, the central region of the tank seems hardly affected by the centrifugal effects, by the sidewalls, by the slight curvature of the free surface, or by non-Boussinesq effects.

In some sense, our experiment follows a strategy similar to Barker et al.~\cite{Barker} while proposing a situation that can be realized in the laboratory. It thus comes as a surprise that the heat transport efficiency measured in the present experiment is significantly greater than the one reported in the idealized numerical setups of Barker et al.~\cite{Barker}, Stellmach et al.~\cite{Stellmach} and Julien et al.~\cite{Julien2012} in Cartesian geometry, and in Gastine et al.~\cite{Gastine} in spherical geometry: the experimentally measured value of the prefactor ${\cal C}$ is approximately twice as large as the value extracted from Barker et al.~\cite{Barker}, it is six times greater than the value reported in Julien et al.~\cite{Julien2012} and three times larger than the value reported in Gastine et al.~\cite{Gastine} (the latter in spherical geometry). In terms of the prefactor appearing in the scaling-law [\ref{scalingRa}], this translates respectively into an experimentally measured ${\cal C}_\mathrm{Ra}$ that is approximately six times greater than the value extracted from Barker et al.~\cite{Barker}, sixty times greater than the value reported in Julien et al.~\cite{Julien2012}, and twenty times greater than the value reported in Gastine et al.~\cite{Gastine}. This points to an unexpected sensitivity of the heat transport efficiency of rapidly rotating turbulent convection to the precise spatial distribution of heat sources and sinks. We confirmed this conclusion experimentally by doubling the value of the absorption length $\ell/H$, from $\ell/H=0.024$ to $\ell/H=0.048$: this change in the geometry of the heat source leads to an increase in the prefactor ${\cal C}$ by approximately $30\%$, and an approximate doubling of the prefactor ${\cal C}_\mathrm{Ra}$. Beyond the observation of the geostrophic turbulence heat transport scaling-law [\ref{scaling32}], our laboratory setup thus offers a unique experimental opportunity to determine the dependence of the prefactor on the distribution of heat sources and sinks, which greatly varies from planets to stars.

%\bibliographystyle{pnas-new}

%\section*{Acknowledgments}
%
%This research is supported by the European Research Council under grant agreement FLAVE 757239. The numerical study was performed using HPC resources from GENCI-CINES and TGCC (grant 2020-A0082A10803 and grant 2021-A0102A10803). KJ acknowledges support from the National Science Foundation, grant DMS-2009319.

%\section*{Methods}

\matmethods{
\subsection*{Radiative heating and effective uniform cooling}

Within the framework of the Boussinesq approximation, and denoting the temperature and velocity fields as $T$ and ${\bf u}$, respectively, the temperature equation for the radiatively heated fluid reads:

\begin{equation}
\partial_t (\rho C T) + {\bf u} \cdot {\boldsymbol{\nabla} (\rho CT)} =  \rho C \kappa \boldsymbol{\nabla}^2 T + \frac{P}{\ell} e^{-z/\ell} \, , \label{eqheat}
\end{equation}
where the radiative heating term -- the last term on the right-hand side --  results from Beer-Lambert's law. In this expression, $z$ denotes the vertical coordinate measured upwards from the bottom of the tank and $P$ is the total heat flux. %The boundaries are thermally insulating, with rigid bottom and sidewalls and a free surface of fluid below the lid. 
The boundaries are thermally insulating: $\boldsymbol{\nabla} T \cdot {\bf n}  =  0$ at all boundaries, with ${\bf n}$ the unit vector normal to the boundary.
Denoting as $\overline{T}(t)$ the spatial average of the temperature field inside the fluid domain, the spatial average of equation [\ref{eqheat}] yields:
\begin{equation}
\frac{\mathrm{d} \overline{T}(t)}{\mathrm{d}t} = \frac{P}{\rho C H}\left( 1- e^{-H/\ell} \right) \, . \label{eqTbar}
\end{equation}
The spatially averaged temperature increases linearly with time. Once the system reaches a quasi-stationary drifting state, the temperature everywhere inside the tank drifts at a mean rate given by the right-hand side of [\ref{eqTbar}]. We can thus extract the power $P$ from the drift of the timeseries.

Consider now the deviation from the spatial mean, $\theta ({\bf x},t)=T({\bf x},t)-\overline{T}(t)$. We form the equation for $\theta$ by subtracting equation [\ref{eqTbar}] from equation [\ref{eqheat}]$/\rho C$:
\begin{equation}
\partial_t \theta + {\bf u} \cdot {\boldsymbol{\nabla} \theta} =  \kappa \boldsymbol{\nabla}^2 \theta + S(z) \, , \label{eqtheta}
\end{equation}
where the source/sink term $S(z)$ is:
\begin{equation}
S(z) = \frac{P}{\rho C} \left( \frac{1}{\ell} e^{-z/\ell} -\frac{ 1- e^{-H/\ell} }{H}  \right) \, .
\end{equation}
The second term inside the parenthesis is an effective cooling term associated with the secular heating of the body of fluid. It balances the heating term on average over the domain but has a different spatial structure.
Equation [\ref{eqtheta}] is coupled to the rotating Navier-Stokes equation:
\begin{equation}
\partial_t {\bf u} + ({\bf u} \cdot \boldsymbol{\nabla}) {\bf u} +2\Omega {\bf e}_z \times {\bf u}=  -\boldsymbol{\nabla} p + \alpha g \theta \, {\bf e}_z + \nu \boldsymbol{\nabla}^2 {\bf u} \, , \label{eqNS}
\end{equation}
where the generalized pressure term absorbs the contribution from the mean temperature $\overline{T}(t)$ and the centrifugal acceleration has been neglected.

The set of equations [\ref{eqtheta}-\ref{eqNS}] corresponds to the standard equations of rotating Boussinesq convection, with internal heating decreasing exponentially with height and uniform cooling at an equal and opposite rate. The solutions to this set of equations reach a statistically steady state, and the temperature difference $\Delta \theta$ realized by [\ref{eqtheta}-\ref{eqNS}] is equal to the temperature difference $\Delta T$ of the initial setup.

\subsection*{Detailed experimental protocol}

An experimental run consists of the following steps: the tank is filled with $7^o$C water mixed with carbon-black dye to obtain a target value of $\ell$. The tank is set into uniform rotation at a rate $\Omega$. After an initial waiting period, for the fluid to achieve solid body rotation, the 2500W metal-halide spotlight is turned on. Two thermocouples horizontally centered inside the tank give access to the temperature at heights $z=0$ and $z=3H/4$. The corresponding temperature signals are measured by an Arduino microcontroller and transmitted through WIFI to a second Arduino microcontroller, which allows for live monitoring of the signals. An example of timeseries is provided in Figure~\ref{fig:timeseries}. After an initial transient phase the system settles in a quasi-stationary state characterized by a linear drift of the two timeseries at an equal rate (visible for $t \gtrsim 1500$s in Figure~\ref{fig:timeseries}), together with a statistically steady temperature difference between the two probes. The fact that the two timeseries drift at a constant and equal rate is a first indication that thermal losses are negligible. We determine the input heat flux $P$ from the drift rate of the two signals using relation [\ref{eqTbar}]. The time-average of the temperature difference between $z=0$ and $z=3H/4$ yields the temperature drop $\Delta T$. This average is performed over the boxed time interval in Figure~\ref{fig:timeseries}. The dimensionless parameters are computed using the fluid properties evaluated for the mean bottom temperature over that interval. To quantify the error associated with the slow temporal drift of the various fluid properties (diffusivities, thermal expansion, etc) we also compute the various quantities and dimensionless parameters using the first and second halves of the boxed region, denoted respectively as sub-region $\text{I}$ and sub-region $\text{II}$ in Figure~\ref{fig:timeseries}. For each of the two sub-regions, we average the temperature difference over the sub-interval and we compute the dimensionless parameters using the mean bottom temperature inside the sub-interval. The corresponding values are reported in the SI with a subscript $\text{I}$ or $\text{II}$ depending on the sub-interval. Also reported are the initial and final bottom temperatures of the boxed measurement interval, denoted as $T_{start}$ and $T_{end}$, respectively. The error bars in Figures~\ref{fig:rawdata} and \ref{fig:rescaleddata} correspond to the values obtained by restricting attention to a single sub-interval. \cor{When estimating the best-fit exponent $\beta$, we first compute the root-mean-square error on $\log {\cal N}$ over the range of ${\cal R}$ of interest (of the order of $5\%$), before propagating this error into a standard deviation $\sigma_\beta$ for the best-fit exponent $\beta$.}

\begin{figure}
%    \centerline{\includegraphics[width=14 cm]{figures/rescaledNu_vs_controlparam_article.eps} }
    \centerline{\includegraphics[width=9 cm]{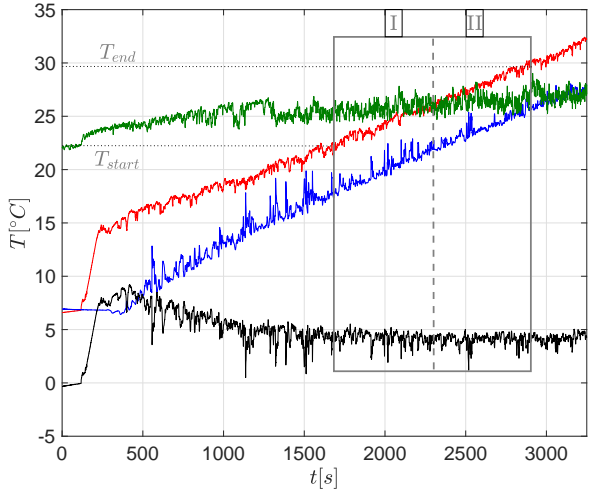} }
   \caption{Raw signals from thermocouples $T_1$ (red, $z=0$) and $T_2$ (blue, $z=3H/4$) as a function of time $t$ for $H=25$cm, $\Omega=30$rpm and $\ell/H=0.048$. Also shown are the instantaneous temperature drop between the two probes (black), and room temperature (green). The solid box indicates the total measurement interval, separated by a dashed-line into two sub-intervals $\text{I}$ and $\text{II}$. \label{fig:timeseries}}
\end{figure}

\subsection*{Direct numerical simulations}

We solve equations [\ref{eqtheta}-\ref{eqNS}] inside a horizontally periodic domain with the pseudo-spectral solver Coral~\cite{Miquel2021}, previously used for non-rotating convective flows~\cite{Miquel2020} and validated against both analytical results~\cite{miquelPRF19} and solutions computed with the Dedalus software~\cite{Dedalus}. The bottom boundary is insulating and no-slip, while the top boundary is insulating and stress-free. Depending on the Ekman number, the horizontal extent $L_\perp$ of the domain is set to $0.4 H$ or $0.5 H$, to account for the variation in the characteristic horizontal scale of the rotating flow. The equations are discretized on a grid containing $(N_x, N_y, N_z)=(441,441,576)$ points, which corresponds to $296$ alias-free Fourier modes in the horizontal directions  and $384$ Chebyshev polynomials along the vertical. The initial condition is chosen as either small amplitude noise or a checkpoint from a previous simulation with smaller supercriticality. We restrict attention to the statistically steady state that arises after the initial transient. We denote as $\tau_\mathrm{meas}$ the duration of integration in this statistically steady state and we focus on the difference between the horizontally averaged temperatures at $z=0$ (bottom boundary) and $z=3H/4$: the time-average of the resulting signal yields the temperature drop from which ${\cal N}$ is inferred. The standard deviation $\sigma$ of the signal and its correlation time $\tau_\mathrm{corr}$ (time-lag of the first zero of the autocovariance function) allow us to estimate the statistical error on ${\cal N}$. Following e.g. Ref.~\cite{Cheng2020}, we compute the number of `effectively independent realizations' $N_\text{eff}=\tau_\mathrm{meas}/\tau_\mathrm{corr}$ before estimating the statistical error $\sigma_{\cal N}$ on the mean temperature drop as $\sigma/\sqrt{N_\text{eff}}$. The resulting error bars are provided in the SI appendix, together with the values of ${\cal N}$ obtained by averaging over only the first or second half of the signal, denoted as ${\cal N}_\text{I}$ and ${\cal N}_\text{II}$, respectively.

}

%\begin{figure}%[tbhp]
%\centering
%\includegraphics[width=.8\linewidth]{frog}
%\caption{Placeholder image of a frog with a long example legend to show justification setting.}
%\label{fig:frog}
%\end{figure}
%
%
%\begin{SCfigure*}[\sidecaptionrelwidth][t]
%\centering
%\includegraphics[width=11.4cm,height=11.4cm]{frog}
%\caption{This legend would be placed at the side of the figure, rather than below it.}\label{fig:side}
%\end{SCfigure*}

\showmatmethods{} % Display the Materials and Methods section

\acknow{This research is supported by the European Research Council under grant agreement FLAVE 757239. The numerical study was performed using HPC resources from GENCI-CINES and TGCC (grant 2020-A0082A10803 and grant 2021-A0102A10803). KJ acknowledges support from the National Science Foundation, grant DMS-2009319.}

\showacknow{} % Display the acknowledgments section

\bibliography{rotating_CONRAD}

\end{document}